\documentclass[aps,letterpaper,twocolumn,preprintnumbers,floatfix,superscriptaddress]{revtex4}

\usepackage{graphicx}
\usepackage{dcolumn}
\usepackage{bm}
\usepackage{epsfig}
\usepackage{gensymb}
\usepackage{mathrsfs}  
\usepackage{amsmath}
\usepackage{amssymb}
\usepackage{graphicx,bm}
\usepackage{slashed} 
\usepackage{dsfont}
\usepackage[makeroom]{cancel}
\usepackage{youngtab}
\usepackage{multirow}
 \usepackage[titletoc]{appendix}

\def\fun#1#2{\lower3.6pt\vbox{\baselineskip0pt\lineskip.9pt
  \ialign{$\mathsurround=0pt#1\hfil##\hfil$\crcr#2\crcr\sim\crcr}}}

\def\lsim{\mathrel{\rlap{\raise 2.5pt \hbox{$<$}}\lower 2.5pt\hbox{$\sim$}}}
\def\gsim{\mathrel{\rlap{\raise 2.5pt \hbox{$>$}}\lower 2.5pt\hbox{$\sim$}}}

\input epsf

\newcommand{\be}{\begin{equation}}
\newcommand{\ee}{\end{equation}}
\newcommand{\bea}{\begin{eqnarray}}
\newcommand{\eea}{\end{eqnarray}}
\newcommand{\mn}{{\mu\nu}}
\newcommand{\B}{\mbox}

\usepackage{color}

\newcommand{\comment}[1]{}

\begin{document}

\title{Emergence of Maximal Symmetry}

\author{Csaba Cs\'aki}
\affiliation{ Department of Physics, LEPP, Cornell University, Ithaca, NY 14853, USA}
\author{Teng Ma}
\affiliation{ 
 CAS Key Laboratory of Theoretical Physics, Institute of Theoretical Physics,
Chinese Academy of Sciences, Beijing 100190, China.}
\author{Jing Shu}
\affiliation{ 
CAS Key Laboratory of Theoretical Physics, Institute of Theoretical Physics,
Chinese Academy of Sciences, Beijing 100190, China.}
\affiliation{School of Physical Sciences, University of Chinese Academy of Sciences, Beijing 100049, P. R. China.}
\affiliation{CAS Center for Excellence in Particle Physics, Beijing 100049, China}
\affiliation{Center for High Energy Physics, Peking University, Beijing 100871, China}
\author{Jiang-Hao Yu}
\affiliation{ 
 CAS Key Laboratory of Theoretical Physics, Institute of Theoretical Physics,
Chinese Academy of Sciences, Beijing 100190, China.}
\affiliation{School of Physical Sciences, University of Chinese Academy of Sciences, Beijing 100049, P. R. China.}

\begin{abstract}
An emergent global symmetry of the composite sector (called maximal symmetry) can soften the ultraviolet behavior of the Higgs potential and also significantly modify its structure. We explain the conditions for the emergence of maximal symmetry as well as its main consequences. We present two simple implementations  and generalize both to N-site as well as full warped extra dimensional models. The gauge symmetry of these models enforces the emergence of maximal symmetry. The corresponding Higgs potentials have unique properties: one case minimizes the tuning while the other allows heavy top partners evading direct LHC bounds. 
\end{abstract}


\maketitle

\section{Introduction}
\label{Sec:intro}

 The discovery of Higgs boson has been a milestone for particle physics~\cite{Aad:2012tfa, Chatrchyan:2012xdj}. However, the potential for such an elementary scalar particle is generically sensitive to physics at extremely high scales, rendering the Higgs potential unstable to quantum corrections. One can impose additional symmetries to eliminate this ultraviolet (UV) sensitivity.    Besides supersymmetry~\cite{Wess:1974tw} or discrete symmetry like twin parity~\cite{Chacko:2005pe} or trigonometric parity~\cite{TP}, one widely considered possibility is a spontaneously broken (approximate) global symmetry, with the Higgs identified as one of the pseudo-Nambu-Goldstone bosons (pNGB) of this symmetry breaking~\cite{Kaplan:1983fs, Georgi:1984af, Dugan:1984hq} (for reviews see~\cite{Contino:2010rs,CHreviews}). Particular implementations   of this global symmetry breaking can forbid the UV divergences of the Higgs potential. Some of the leading ideas along this direction are collective symmetry breaking and little Higgs~\cite{ LH} models, dimensional deconstruction~\cite{ArkaniHamed:2001ca, ArkaniHamed:2001nc}, warped extra dimensions~\cite{Randall:1999ee,Contino:2003ve, Agashe:2004rs}, and the Weinberg sum rule for a composite Higgs~\cite{Marzocca:2012zn}.

Recently a new concept has been proposed as an alternative to these methods mentioned above; the UV divergences of the Higgs potential from the top sector are absent because of ``maximal symmetry"~\cite{Csaki:2017cep}. The structure of the low energy effective Lagrangian differs from the generic case: maximal symmetry forbids Higgs corrections for the effective kinetic terms of the top quark which source the UV divergence and are often the leading sources for the quadratic term in the Higgs potential. Although maximal symmetry is simple, elegant and can have many model building applications, the exact nature of maximal symmetry and its emergence in the low-energy effective action have remained somewhat mysterious. One would like to understand what exactly the origin of this symmetry is and how to systematically realize it in the low energy effective Lagrangian by integrating out the heavy composite fields or some bulk KK modes in extra dimensions.

In this paper, we show that the origin of maximal symmetry is actually very simple: it is simply an enhanced global symmetry of the composite sector! Generically composite Higgs models are based on a coset $G/H$ corresponding to the $G\to H$ symmetry breaking pattern and the composite sector only has an $H$ symmetry. Whenever $H$ is  enhanced to $G$ we will obtain a maximally symmetric model.  After explaining the basic principles behind maximal symmetry we demonstrate them by  constructing the simplest two site models. We show that maximal symmetry can be easily enforced by the gauge symmetries   of the model,  indicating that maximal symmetry is secretly a remnant of some of the gauge symmetries broken at higher energies.  Next we generalize this simple models to the N-sites. In these models the fermions at the intermediate sites automatically have a global $G$ symmetry, and the key point of realizing maximal symmetry is to preserve a global symmetry $G$ or $G'$ other than $H$ for the fermions  at the last site. Using  this observation, we also build simple warped extra dimensional models realizing the two implementations of maximal symmetry. Note that the second implementation  (``minimal maximal symmetry") is a brand new setup that has never been discussed before.  
We finally briefly discuss the structure of the Higgs potential and find that a light Higgs can be obtained without light top partners in the model with minimal maximal symmetry.

\section{Maximal Symmetry} 
\label{sec:MS}

Maximal symmetry is the enhancement of the spontaneously broken $G/H$ symmetry back to the full $G$ in some sector of the composites. Generically composites form representations of the unbroken group $H$, and the original $G$ symmetry is not implemented in the composite sector. In the most general situation the the composites don't even have to fill out a complete $G$ representation. There may however be sectors of composite fields (for example the fermionic top partners) where the composites themselves still form a complete $G$ multiplet, hence the composite sector may have an emergent enhanced $G$ global symmetry. This enhanced global symmetry is maximal symmetry, which can play an important role in the structure of the induced Higgs potential. 

For concreteness we will consider the $G=SO(5)$ and $H=SO(4)$ symmetry breaking pattern corresponding to the minimal choice that incorporates a custodial symmetry for the SM. The Higgs field is contained in the non-linear sigma field $U$ which transforms as~\cite{Coleman:1969sm, Callan:1969sn}

\begin{equation}
U\to g U h^\dagger
\end{equation}
where $g \in G$ is an element of the linearly realized full $G$ symmetry while $h\in H$ is the non-linearly realized shift-symmetry.  Thus the $U$ field can also be interpreted as the field connecting $SO(5)$ symmetry of the elementary sector with the spontaneously broken $SO(5)$ symmetry of the composite sector: it transforms under an $SO(5)_{el}\times SO(5)_{co}$ symmetry, where the composite sector breaks $SO(5)_{co}$ to $SO(4)$. The special case of maximal symmetry is when some remnant of the full $SO(5)_{el}\times SO(5)_{co}$ is left over providing additional protection for the Higgs potential and softening its UV behavior. If either $SO(5)_{el}$ or $SO(5)_{co}$ is unbroken there will simply be no Higgs potential: either of these symmetries is sufficient to fully protect the pNGB's from acquiring any potential. We thus need to break both $SO(5)$'s but in such a manner that some remnant bigger than $SO(4)$ is left over. There are two simple options emerging, depending on the embedding of the SM fermions into the global symmetries. 

\begin{figure}
\begin{center}
\includegraphics[width=0.5\columnwidth]{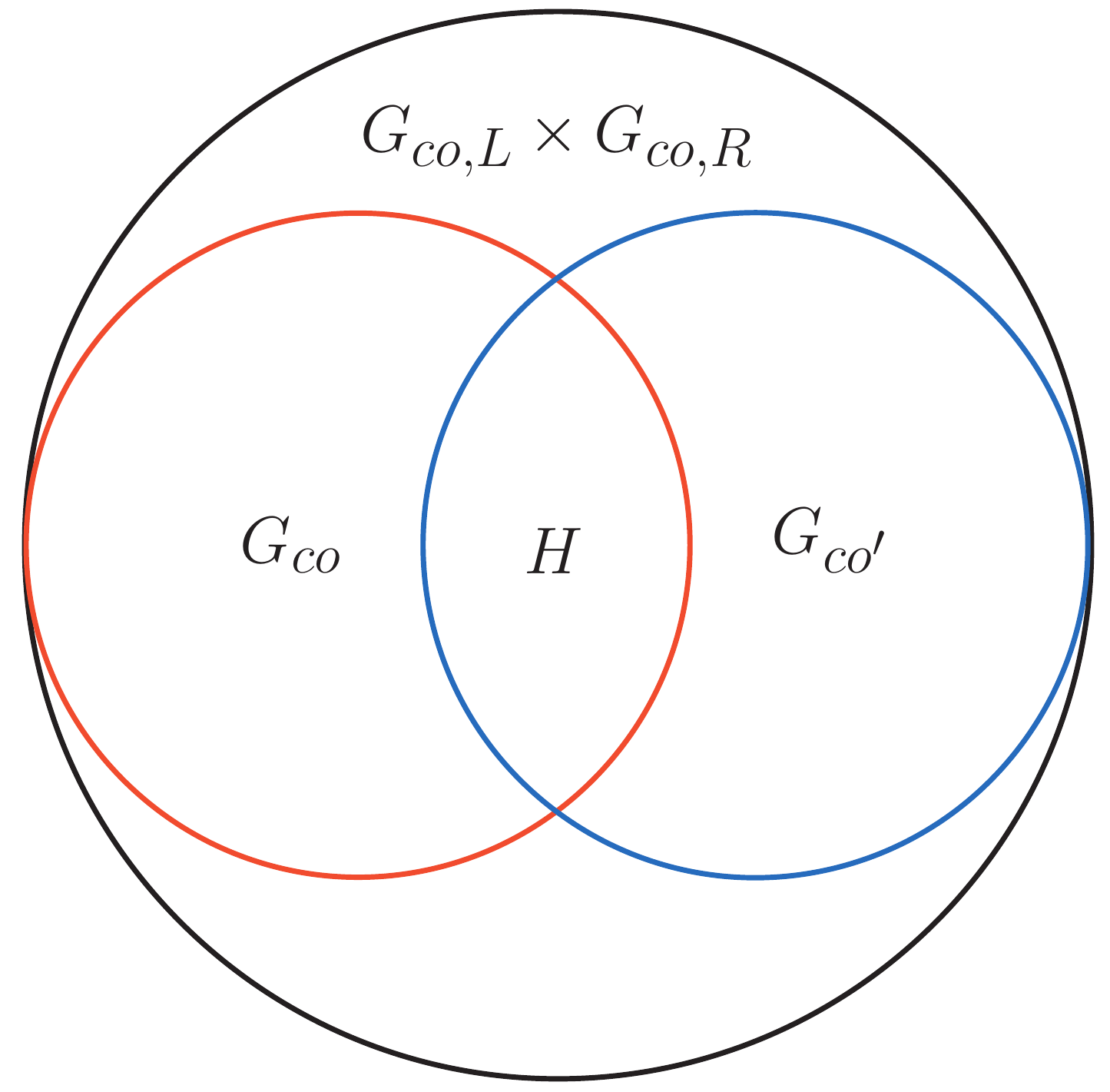}
\end{center}
\caption{Sketch of the pattern of symmetries leading to maximal symmetry in the composite sector.} \label{fig:maximal_symmetry}
\end{figure} 

1. Both the left handed top doublet $q_L$ and the right handed top $t_R$ are embedded into $SO(5)_{el}$. This is the standard assumption, corresponding to the SM fermions being mainly elementary. The embedding of these fields into incomplete $SO(5)_{el}$ multiplets breaks the elementary symmetry, but does not say anything about the structure of the composite sector. The enhancement of the global symmetries will depend entirely on the structure of the composite fields. To achieve our goal we need to preserve an $SO(5)$ symmetry that does not coincide with the original $SO(5)_{co}$. The original proposal of maximal symmetry is exactly that: an $SO(5)_{co'}$ symmetry that appears in the composite sector, where the $SO(5)_{co'}$ is not identical to $SO(5)_{co}$, see Fig.\ref{fig:maximal_symmetry}.   

2. The second option is when $q_L$ is embedded in the elementary sector, but $t_R$ in the composite sector. Since $t_R$ is an $SU(2)_L$ singlet this can be easily achieved by simply making $t_R$ a singlet under $SO(4)_{co}$. In this case already the embedding of $q_L$ and $t_R$ will have the right symmetry breaking pattern of $SO(5)_{el}\times SO(5)_{co}$ to ensure that a Higgs potential will be generated. If the remaining composites maintain any form of $SO(5)$ symmetry (which now could also coincide with the original $SO(5)_{co}$) a softening of the UV behavior of the Higgs potential is expected. We call this new possibility the minimal realization of maximal symmetry.  

 \section{Implications of Maximal Symmetry\label{sec:Implications}}

We have argued that maximal symmetry is an emergent accidental symmetry in the composite sector. Next we show how such a symmetry will potentially soften the UV behavior of the Higgs potential. Consider first the case considered in~\cite{Csaki:2017cep} when both $q_L$ and $t_R$ are embedded in the elementary sector. For concreteness we assume that they are both in fundamentals of $SO(5)_{el}$ transforming as $\Psi_{q_L}\to g \Psi_{q_L}$ and $\Psi_{t_R}\to g \Psi_{t_R}$. We also assume that the coset $G/H$ is   a so-called  ``symmetric space" (which means that there exists a Higgs-parity operator $V$). In this case one can always construct the linearly realized pNGB matrix $\Sigma' =UVU^\dagger = U^2 V$, which transforms linearly under the full set of global symmetries $\Sigma' \to g \Sigma' g^\dagger$. By integrating out the composite sector one can always find the effective Lagrangian for the elementary fields. Since the composite sector is fully integrated out, its effect will show up only via some form factors, and the effective Lagrangian can be constrained by only considering the elementary symmetries. This will restrict the form to be~\cite{Csaki:2017cep}
\bea
\label{eq:gene_maximal}
\mathcal{L}_{\mbox{eff}} 
  &=& \bar{\Psi}_{q_L} \slashed p(\Pi_0 ^L(p) +\Pi_1 ^L(p) \Sigma') \Psi_{q_L} - \bar{\Psi}_{q_L} M_1 ^t(p) \Sigma' \Psi_{t_R} \nonumber \\
  &+& \bar{\Psi}_{t_R} \slashed p (\Pi_0 ^{R}(p) +\Pi_1 ^{R}(p) \Sigma' )\Psi_{ t_R}  +h.c.  \ ,
\eea
where the form factors $\Pi_{0,1}^{L/R}(p)$ and $M_1^t$ encode the effects of the composite sector. A global symmetry in the composite sector will manifest itself in some special relation among the form factors, in most cases some of the form factors are simply vanishing due to the symmetry. Most interesting is the example where $\Pi_1^{L,R}=0$ due to the symmetry in the composite sector, the example considered in~\cite{Csaki:2017cep}, while $M_1^t \neq 0$. Note that $\Pi_1^{L,R}=0$ automatically ensures the finiteness of the Higgs potential. How can one use composite global symmetries to forbid $\Pi_1$ but allow $M_1^t$ (which is necessary to generate any Higgs potential)? The method used in~\cite{Csaki:2017cep} was to introduce an $SO(5)_{co, L}\times SO(5)_{co,R}$ chiral global symmetry in the composite sector and ensure that the composite mass term breaks this chiral global symmetry to $SO(5)_{co'}$ defined by $g_{co,L} V g_{co,R}^\dagger = V$ for $g_{co,L}\in SO(5)_{co,L}$ and $g_{co,R} \in SO(5)_{co,R}$. One can then trace back the action of the composite symmetries in the Lagrangian (\ref{eq:gene_maximal}) by noting that the dressed elementary fields $U^\dagger \Psi_{q_L, t_R}$ transform under the composite symmetries. For the chirally enhanced composite global symmetries $U^\dagger \Psi_{q_L, t_R} \to g_{co, L,R} U^\dagger \Psi_{q_L, t_R}$. The $\bar{\Psi}_{q_L} \slashed{p} \Sigma' \Psi_{q_L}$ can be rewritten in terms of the dressed fields as 
$(\bar{\Psi}_{q_L} U) V (U^\dagger \Psi_{q_L})$ and is not invariant under the $SO(5)_{co'}$ symmetry. The structure of the $M_1^t$ term however exactly coincides with the symmetry breaking pattern of the chiral composite symmetries $\bar{\Psi}_{q_L} \Sigma' \Psi_{t_R} = (\bar{\Psi}_{q_L} U) V (U^\dagger \Psi_{t_R})$ which is invariant for transformations obeying $g_{co,L} V g_{co,R}^\dagger =V$. 

Let us now consider the second possibility, not discussed so far in the literature, which we will refer to as the ``minimal maximal symmetry".  In this case
$q_L$ is still embedded in the elementary sector, however $t_R$  is now assumed to be transforming under the global symmetries of the composite sector. Note that this does not necessarily imply that $t_R$ is a composite itself. Since $t_R$ is an $SU(2)_L$ singlet, it can for example easily mix with a singlet from the composite sector {\it without} being dressed by the $U$ field. Such a mixing would imply that $t_R$ will be transforming under the composite global symmetries rather than the elementary ones.   For simplicity we will again assume that $q_L$ and $t_R$ are embedded into fundamental representations of $SO(5)_{el}$ and $SO(5)_{co}$ respectively, transforming as $\Psi_{q_L} \to g_{el} \Psi_{q_L}$  and $\Psi_{t_R} \to g_{co} \Psi_{t_R}$ with $g_{el} \in SO(5)_{el}$ and $g_{co}\in SO(5)_{co}$. Since the composite sector must be $H \equiv SO(4) \subset SO(5)_{co}$ invariant, $\Psi_{t_R}$ should be a full $H$ representation to keep $H$ unbroken. After integrating out the composite sector the general form of  the effective Lagrangian invariant under the  $SO(5)_{el}$ global symmetry can be written as 
 \bea 
\label{eq:mini_maximal}
\mathcal{L}_{\mbox{eff}} 
    &=& \bar{\Psi}_{q_L} \slashed p(\Pi_0 ^L(p) +\Pi_1 ^L(p) \Sigma') \Psi_{q_L} +\bar{\Psi}_{t_R} \slashed p \Pi_0 ^{R}(p) \Psi_{ t_R} \nonumber \\
  &+& \bar{\Psi}_{q_L} M_1 ^t(p)U\Psi_{t_R} +h.c. 
\eea                                                     
differing slightly from (\ref{eq:gene_maximal}): the form factor $\Pi_1^R$ is automatically vanishing, while the $M_1^t$ mass term has a $U$ insertion connecting the elementary and composite sectors, rather than the $\Sigma'$. 
Following the discussion in the first case, if any $SO(5)$ subgroup of the chirally enhanced composite global symmetries, defined as  $U^\dagger \Psi_{q_L} \to g_{co, L} U^\dagger \Psi_{q_L}$ and $\Psi_{t_R} \to g_{co, R} \Psi_{t_R}$, is unbroken,  the form factor $\Pi_1^L$  will again be  forbidden. However the $M_1^t$ term is automatically invariant under this symmetry and will be allowed. Note that in some sense this scenario is even more powerful than the traditional implementation of maximal symmetry. Usually one needs to choose an alignment for the composite mass terms to point exactly in the $SO(5)_{co'}$ direction, one which will also leave the $M_1^t$ term in the effective action invariant. For the minimal maximal symmetry however {\it any} $SO(5)$ subgroup of the chiral global symmetries is sufficient - the modified $M_1^t$ term  will always be left invariant. However the embedding of the $q_L$ and $t_R$ into $\Psi_{q_L}$ and $\Psi_{t_R}$ will now explicitly break both the elementary and the composite global symmetries, and a Higgs potential will be generated.

\section{Simplest Model for Finite EWSB: 2-site Model with Maximal Symmetry}
\label{sec:2_site}
In this section we present the simplest concrete examples of models with maximal symmetry. These also represent the simplest realistic finite EWSB models. One of the main takeaways from these models is that gauge symmetry can be used to enforce the relations needed for the appearance of maximal symmetry, and no special tuning or coincidence of parameters is needed to achieve the maximally symmetric limit. We present the two-site models corresponding to both implementations of maximal symmetry explained in detail in our general discussion above.
 Later we will show how to generalize them  to N sites as well as to full extra dimensional constructions.

\subsection{The Minimal Model for a Maximally Symmetric Composite Higgs}
\label{sec:2_site_1}

\begin{figure}
\begin{center}
\includegraphics[width=0.45\columnwidth]{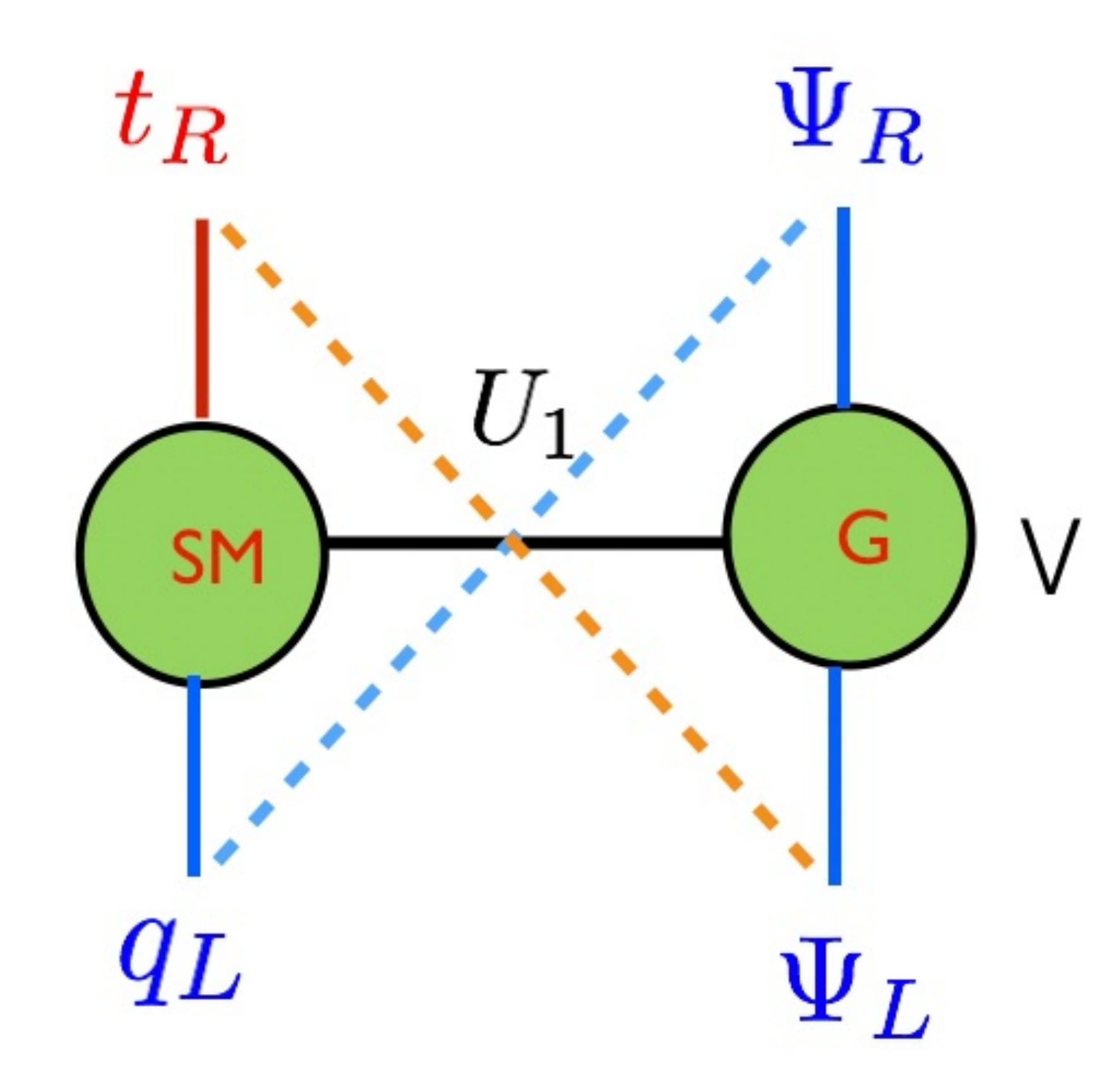}
\includegraphics[width=0.45\columnwidth]{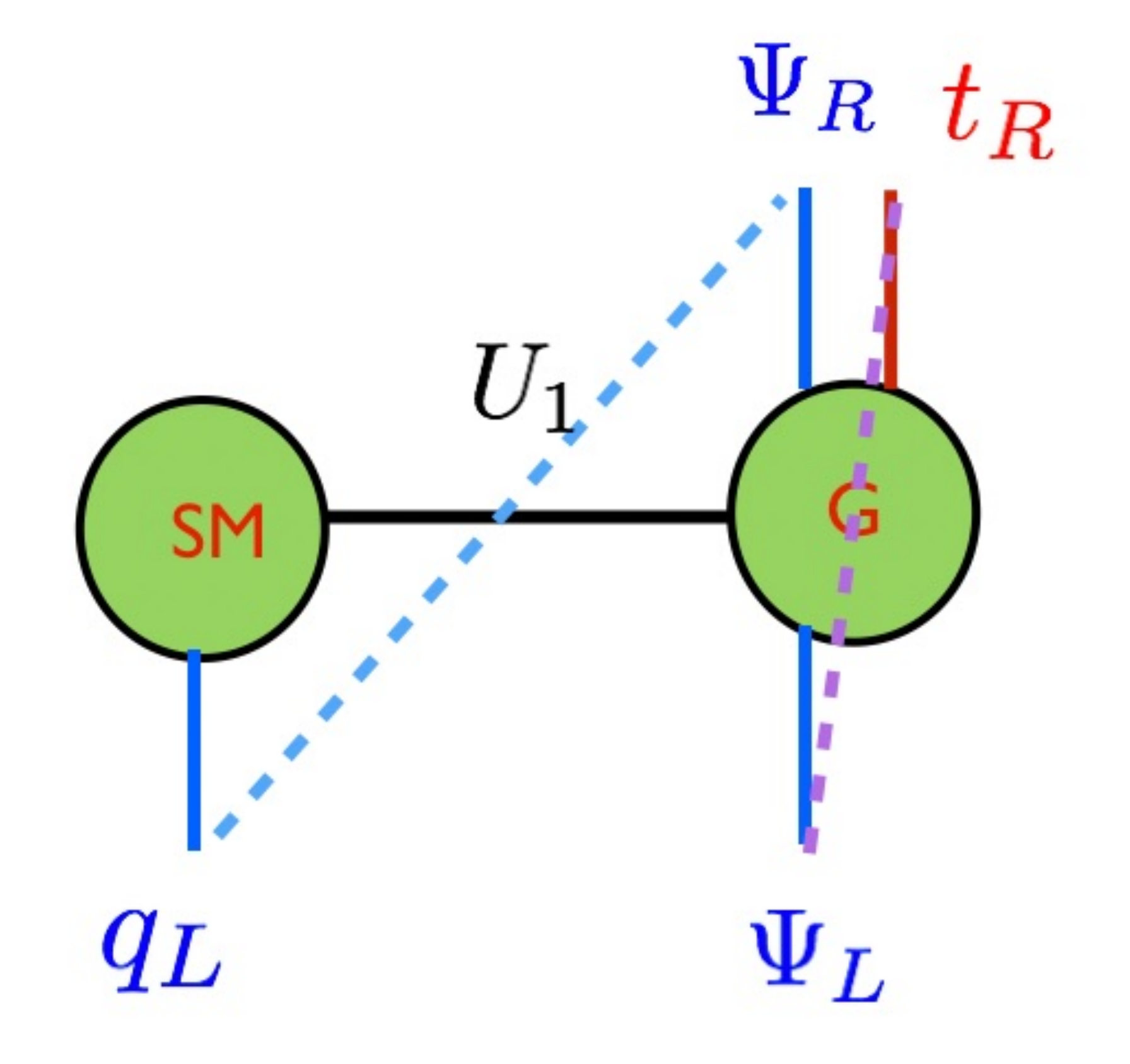}
\end{center}
\caption{Moose diagrams for the two-site models. Left: ordinary maximal symmetry; Right: minimal maximal symmetry.} \label{fig:two_site}
\end{figure} 

 The two-site model can realize maximal symmetry in an extremely simple way.  The appearance of the maximal symmetry will be a consequence of the gauge symmetries of the model. For concreteness, we take a two-site model realizing the coset $SO(5)/SO(4)$~\cite{Agashe:2004rs}  as an example. We will have a global $SO(5)$ symmetry at both sites,  hence the full global symmetry of the moose is $SO(5)_1 \times SO(5)_2$.  The link field $U_1$ connecting these two sites is in the bi-fundamental representation of the global symmetry, breaking it to the diagonal subgroup $SO(5)_V$.  We gauge the $SU(2)_L \times U(1)_Y$ subgroup of $SO(5)_1$ at the first site, which will correspond to the usual  EW symmetry, while at the second site the entire $SO(5)_2$ is fully gauged, as shown in the left panel in Fig.~\ref{fig:two_site}.\footnote{The gauging of the $SO(5)_2$ is the main difference compared to the ordinary deconstruction model discussed in appendix~\ref{app:MCHM}.}   This gauged $SO(5)_2$ will be at the heart of the appearance of the maximal symmetry. In order for the moose to realize the $SO(5)/SO(4)$ coset, the gauge symmetry at the last site should be broken to $SO(4)$. To achieve this we will use a VEV in the symmetric representation of $SO(5)_2$. The choice of a symmetric VEV is the main difference in the construction of the gauge sector of our 2-site model compared to the traditional deconstructed models of minimal composite Higgs  where the breaking is usually achieved via the 5-dimensional vector representation of $SO(5)$. The reason 
for the choice of a symmetric VEV
is that this will be coinciding with the Higgs parity operator $V={\rm diag} (1,1,1,1,-1)$, allowing the possible appearance of maximal symmetry. The linear pNGB field $\Sigma$ corresponding to this breaking can be parametrized as $\Sigma =U^{\prime} V U^{\prime \dagger}$, where $U^\prime$ is non-linear sigma field of coset space $SO(5)_2/SO(4)$. The $SO(5)_1$ and $SO(5)_2$ global symmetries (below the breaking scale of the gauged $SO(5)_2$) can be identified with the 
  $SO(5)_{el}$  and  $SO(5)_{co}$  symmetries of the general discussion in the Section~\ref{sec:MS}. The breaking of the gauge symmetries will eat some of the pNGB's such that in the end we are left with a single set of pNGBs corresponding to the $SO(5)/SO(4)$ coset.
These uneaten NGBs can be described by the linear sigma field $\Sigma^\prime =UVU^\dagger$ with $U=U_1 U^\prime$. 

For the fermion sector of the model we introduce the LH top doublet and the RH top singlet at the first site, while at the second site we introduce a complete Dirac fermion $\Psi$ in the fundamental representation of the entire $SO(5)_2$ - this is required by the fact that we have gauged $SO(5)_2$. This is the point where the gauging of the global symmetry will require that the composite fermions fill out a complete $SO(5)_2$ multiplet, a prerequisite for the emergence of maximal symmetry.  This fermion multiplet will have an enhanced  $SO(5)_{2L} \times SO(5)_{2R}$ chiral global symmetry in the limit when it is massless, and its bare Dirac mass  breaks this chiral global symmetry to $SO(5)_2$. On the other hand its Yukawa coupling to $\Sigma$ breaks it to $SO(5)_{2^\prime}$: a differently oriented $SO(5)$ subgroup of the chiral global symmetries which keeps the VEV $V$ invariant $g_L V g_R^\dagger = V$. As discussed in the previous section, the appearance of such a global $SO(5)$ symmetry differing from the Higgs shift symmetry is necessary to obtain a maximal symmetric model. We can see that $SO(5)_{2^\prime}$ is the only candidate for such a symmetry. However the Dirac mass for $\Psi$ would break this symmetry, hence we have to assume that $\Psi$ has no bare mass but obtains a mass only from the $SO(5)/SO(4)$ breaking.  This is a technically natural assumption that could also be enforced by a discrete $Z_2$ or $Z_3$ symmetry. The SM fermions $t_L, t_R, b_L$ should be embedded in the fundamental representation of $SO(5)_1$ so that they can mix with $\Psi$:
 \bea
\Psi_{q_L}=\frac{1}{\sqrt{2}} \left( \begin{array}{c}
i b_L \\ 
b_L \\ 
i t_L \\ 
- t_L \\ 
0
\end{array} \right)  \quad  \Psi_{t_R}=\left( \begin{array}{c}
0 \\ 
0 \\ 
0 \\ 
0 \\ 
t_R
\end{array} \right)   . 
\eea    
The most general interactions for these fermions invariant under the gauge symmetries and with a vanishing bare mass are given by
\bea
\mathcal{L}_f &=&\bar{q}_L i\slashed D q_L +   \bar{\Psi} i\slashed D \Psi + \bar{t}_R i\slashed D t_R \nonumber \\
&-&\epsilon_L \bar{\Psi}_{q_L} U_1 \Psi_{R}  - M  \bar{\Psi}_{L} \Sigma \Psi_{R} - \epsilon_R \bar{\Psi}_L U_1^\dagger \Psi_{t_R} +h.c.   \nonumber \\
\eea
 The global symmetry of the composite fermions in the absence of the Yukawa terms is enlarged to a chiral global $SO(5)_{co L}\times SO(5)_{co R}$. However the Yukawa coupling of $\Psi$ to $\Sigma$ breaks this enlarged chiral global symmetry to the maximal symmetry $SO(5)_{co^\prime}$ which keeps the VEV $V$ invariant $g_L V g_R^\dagger = V$. The appearance of this enhanced global symmetry in the fermion sector will ensure the vanishing of the $\Pi_1^{L,R}$ form factors
 in the effective Lagrangian in Eq.~(\ref{eq:gene_maximal}) . This can also be explicitly seen via the following analysis. If we turn off any one of the three Yukawa couplings $\epsilon_{L,R}$ or $M$, the Higgs shift symmetry is restored. Thus the Higgs potential must be proportional to the product of these three couplings, which indicates that the only dependence on the Higgs field will be via the top Yukawa coupling, as expected in models with maximal symmetry. Since power counting tells us that the leading contribution to the Higgs potential should be proportional to the square of the effective top Yukawa coupling,  the Higgs potential will be finite at one-loop for this minimal model.

  Note that eventually this model can  be easily  promoted to full extra dimensional theories by identifying the first site with a UV brane and the second site with an IR brane.

\subsection{2-site Model with Minimal Maximal Symmetry}
\label{sec:2_site_2}
The two site model can also easily realize the minimal implementation of maximal symmetry described in Sec.~\ref{sec:MS}. For this we can choose the same basic model with two sites and $SO(5)_1 \times SO(5)_2$ broken to $SO(5)_V$, and the $SU(2)_L\times U(1)_Y$ gauged on the first site. The main difference will be that the singlet top $t_R$ will be introduced at the second site $SO(5)_2$ as a gauge singlet, as shown in the right panel in  Fig.\ref{fig:two_site}. In addition the $SO(5)_2/SO(4)$ breaking will this time be via a VEV $\mathcal{V}=(0,0,0,0,1)$ in the vector $5$ of $SO(5)_2$ with its corresponding sigma field $\mathcal{H}^\prime =U^\prime \mathcal{V}$. The uneaten NGBs are still in the coset $SO(5)_1/SO(4)$ which can again be described by $\mathcal{H} =U\mathcal{V}$ and $U=U_1 U'$. The bare mass term of Dirac fermion $\Psi$ can be introduced to breaks its chiral symmetry to $SO(5)_2$ as maximal symmetry. The most general Lagrangian then is
\bea
\mathcal{L}_f &=&\bar{q}_L i\slashed D q_L +   \bar{\Psi} i\slashed D \Psi + \bar{t}_R i\slashed D t_R \nonumber \\
&-&\epsilon_L \bar{\Psi}_{q_L} U_1 \Psi_{R}  - M  \bar{\Psi}_{L} \Psi_{R} - \epsilon_R \bar{\Psi}_L \mathcal{H}^\prime t_R +h.c.   
\eea
Following the same analysis, the Higgs potential is still dependent on the product of mixing Yukawa couplings and Dirac mass.  Again we can see that because there is a $SO(5)$ global symmetry in $\Psi$ sector the effective kinetic terms of SM field are independent on Higgs field.

\section{N-site Model for Maximal Symmetry}

In the previous section we explained how maximal symmetry can be realized in a simple two site model and how it can be used for a realistic model of EWSB. Here we generalize this to an N-site model with maximal symmetry  whose topology is just a 1 dimensional interval. This will also allow us to find a simple  extra dimensional realization of the model in the continuum limit. We also explain how the N-site model with maximal symmetry  differs from the usual deconstructed composite Higgs models (CHMs)~\cite{Panico:2011pw,DeCurtis:2011yx}.

The N-site model can be obtained  from the two site model presented above by inserting $N-2$ intermediate sites, which corresponds to the bulk in the extra dimensional theory. The global symmetry $G$ at each intermediate site is fully gauged, corresponding to the bulk gauge symmetry. This gauge symmetry  guarantees that a Dirac fermion at this site will have a global $G$ symmetry, which is the diagonal subgroup of its enhanced chiral symmetry broken by a bare mass term. Thus one can obtain a maximally symmetric N-site model as long as the last site leaves this global $G$ symmetry unbroken. For the last site we can just use the same structures as we used for the two site models presented above.

\begin{figure}
\begin{center}
\includegraphics[width=\columnwidth]{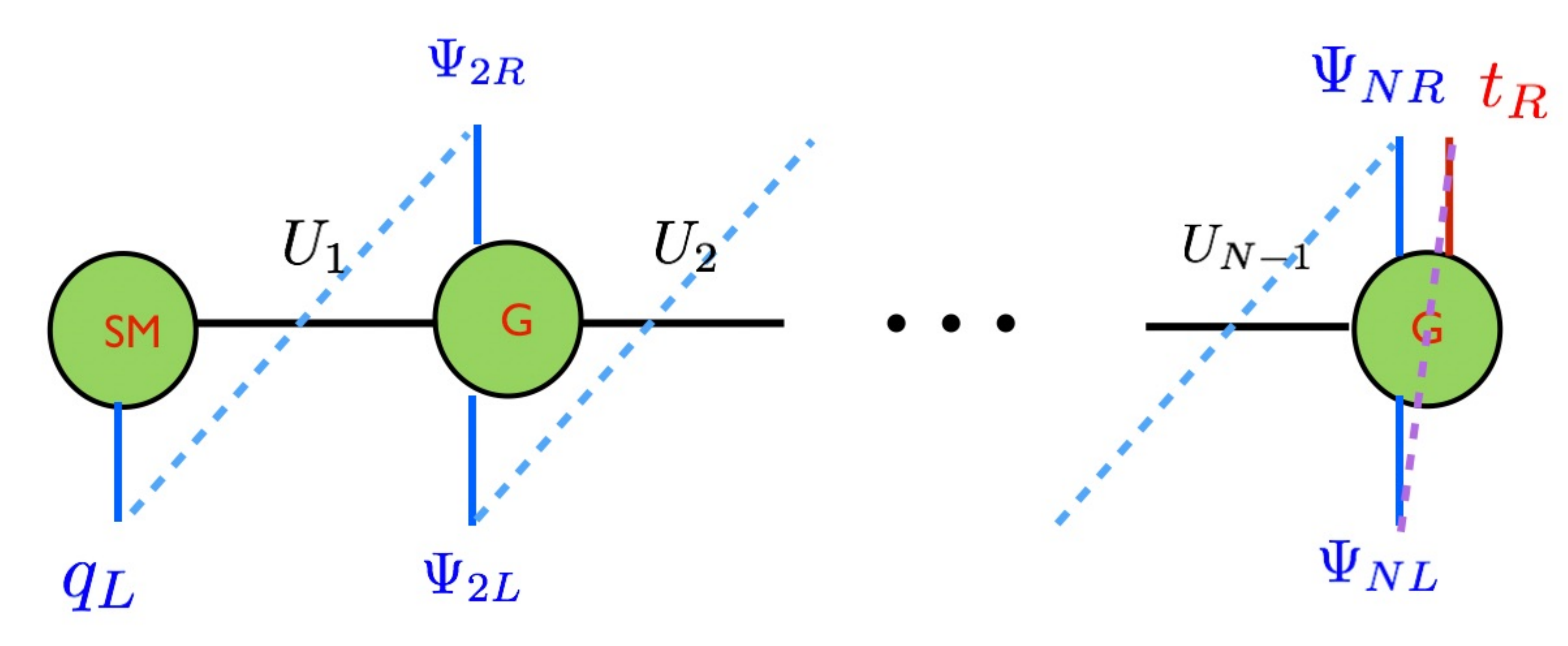}
\end{center}
\caption{Moose diagrams for the N-site model with minimal maximal symmetry.  \label{fig:nsite1}}
\end{figure}

We will briefly present the explicit form of the N-site model corresponding to the $SO(5)/SO(4)$ minimal composite Higgs model (MCHM)~\cite{Agashe:2004rs}, which is a simply direct generalization of the 2-site model.  First we discuss the case of minimal maximal symmetry  with a chiral $\Psi_{Q_L}$ on the first site where only the $SU(2)\times U(1)_Y$ subgroup of $SO(5)$ is gauged, while the chiral singlet $t_R$ is on the last $N^{th}$ site, 
where the $SO(5)$ gauge symmetry is broken to $SO(4)$ by the scalar $\mathcal{H}^\prime$.
  The bulk of the moose is obtained by introducing the Dirac fermions $\Psi_{i}, i=2,\ldots , N$ in the fundamental representation of $SO(5)$ living at the $ i^{th}$ site and with masses $M_{i}$. These will be the composite top partners forming a KK tower of top partners. We also introduce the  the link fields $U_i$, $ i=1,\ldots , N-1$ which will connect the neighboring fermions. Similar to the case of the 2-site model we  find that the intermediate fermions interacting with the SM top sector preserve the global $G$ symmetry. The most general Lagrangian realizing maximal symmetry is then given by 
\bea
\mathcal{L}_f &=& \bar{q}_L i\slashed D q_L + \sum_{i=2}^N  \bar{\Psi}_i i\slashed D_i \Psi_i + \bar{t}_R i\slashed D t_R \nonumber \\
&-&\epsilon_1 \bar{\Psi}_{q_L} U_1 \Psi_{2R}  -\sum_{j=2}^{N-1}  \epsilon_j\bar{\Psi}_{jL}   U_j \Psi_{j+1R}    \nonumber \\
&-&  \sum_{i=2}^N M_i  \bar{\Psi}_{iL} \Psi_{iR} - \epsilon_N \bar{\Psi}_{N L} \mathcal{H}^\prime t_R +h.c.   
\eea
where $D_i \Psi_i = \partial_\mu \Psi_i-i \rho_i \Psi_i$ and $\rho_i$ is the gauge boson at the $\bf i$th site. An illustration of this model can be seen in Fig.~\ref{fig:nsite1}.
 The uneaten Goldstone bosons corresponding to the $SO(5)/SO(4)$ coset  is contained in the product  operator $U=\prod_{i=1}^{N-1}  U_i U^\prime$ which in the continuum limit clearly corresponds to the Wilson line connecting the endpoints of the $5D$ interval. $U$ transforms under the  $SO(5)_{L_1} \times SO(5)_{R_{N}}$ global symmetry as 
\bea
U \to g_{L_1} U g_{R_{N}}^\dagger, 
\eea
which  corresponds to the enlarged $SO(5)_{el} \times SO(5)_{co}$ shift symmetry of our general discussion. This enlarged global symmetry is explicitly broken by the embedding of $q_L$ into $\Psi_{q_L}$ and $t_R$ into $\Psi_{t_R} \equiv \mathcal{V}t_R$.    As explained before, such global symmetry breaking pattern will forbid the Higgs dependent $\Pi_1$ form factors in the kinetic terms but allow the $M_1$ form factor responsible for the top Yukawa coupling, and will generate the finite Higgs potential. 
 
We can see the emergence of maximal symmetry and the specific form of the effective action in  Eq.~(\ref{eq:gene_maximal}) in more detail by simply integrating out the $\Psi_i$ Dirac fermions site by site (corresponding to integrating out the bulk in the continuum limit). Since the bulk Dirac fermion at each site has an unbroken global $SO(5)_{V_i}$ symmetry we can always redefine $\Psi_i$ 
\bea \label{eq:redefinition}
\Psi_{iL} \to \prod_{j=1}^i U_j \Psi_{iL} \quad \Psi_{iR} \to \prod_{j=1}^i U_j \Psi_{iR} \quad  2 \le i \le N \ .
\eea    
In this redefined basis the kinetic term of the SM fermions and the $\Psi_i$ mass terms will be independent of the NGBs. After integrating out all the bulk fermions $\Psi_i$, the NGBs only remain in the link between $\Psi_{q_L}$ and $\Psi_{t_R} \equiv \mathcal{V}t_R$ as $\Psi_{q_L} U  \Psi_{t_R}$. 
Therefore we find that the effective Lagrangian for $\Psi_{q_L}$ and $\Psi_{t_R}$ is invariant under $SO(5)_{L_1} \times SO(5)_{R_N}$  with the form exactly as predicted in  Eq.~(\ref{eq:mini_maximal})
\bea
\mathcal{L}_f  &=& \bar{\Psi}_{q_L} \slashed p \Pi_0^L(p) \Psi_{q_L} + \bar{\Psi}_{t_R} \slashed p \Pi_0^R(p) \Psi_{t_R}  \nonumber \\ 
&-&\Big(  M_1(p) \bar{\Psi}_{q_L} U  \Psi_{t_R} +h.c.\Big) .  
\eea 
We can see that the terms for $\Psi_{q_L}$ and $\Psi_{t_R}$ have the $SO(5)_{L_1}$ and $SO(5)_{R_N}$ global symmetry and the mass term breaks it into the diagonal part giving the minimal implementation of maximal symmetry.

\begin{figure}
\begin{center}
\includegraphics[width=\columnwidth]{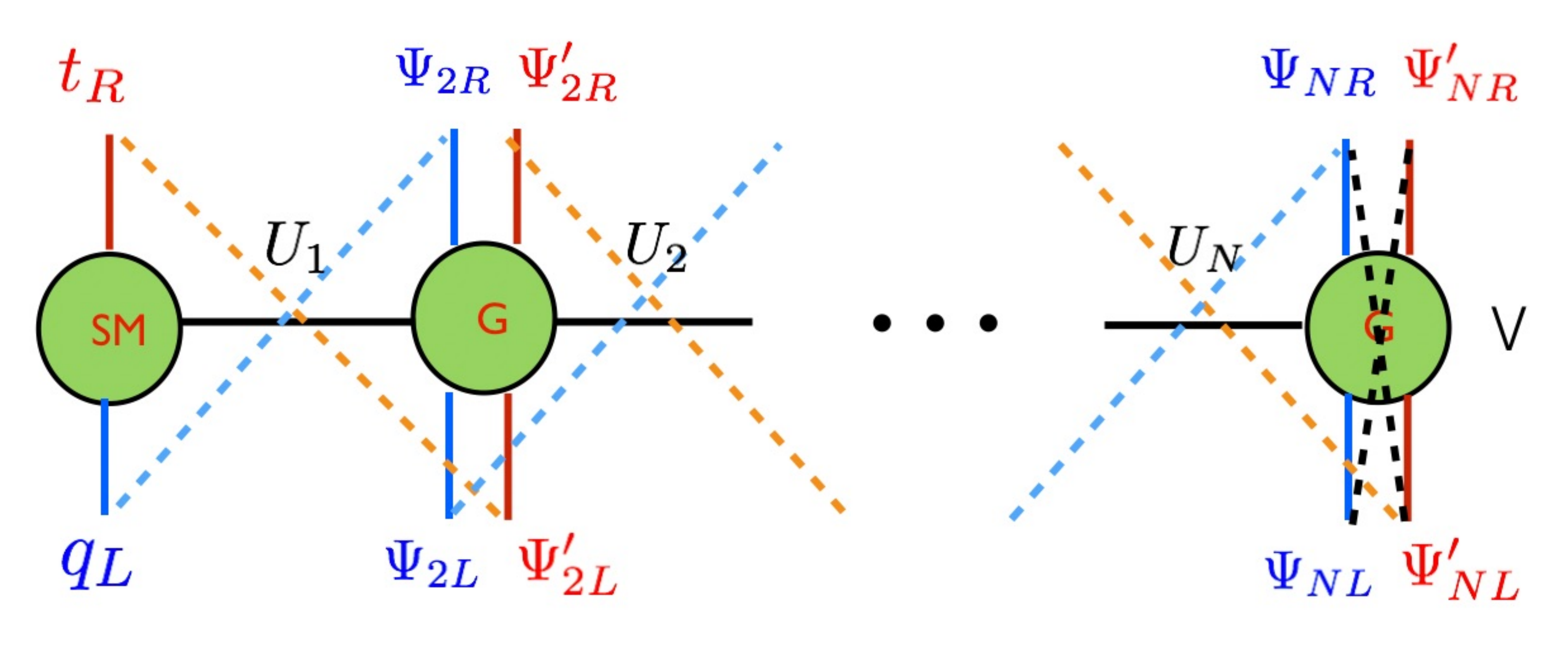}
\end{center}
\caption{Moose diagrams for the N-site model with ordinary maximal symmetry.} \label{fig:nsite2}
\end{figure}

Next we consider the N-site model where both the top doublet and singlet live at the first site, similar to the two site model discussed in section~\ref{sec:2_site_1}. In this case there will be a separate bulk fermion $\Psi$ coupling to $q_L$ and another $\Psi'$ coupling to $t_R$, both in the $\bf 5$ representation of $SO(5)$.  In the continuum limit the top doublet and singlet will be the zero modes of the bulk fermions $\Psi$ and $\Psi^\prime$. We are adding the second bulk fermion $\Psi'$ (unlike in the corresponding 2-site model) to obtain a theory that has a simple extra dimensional interpretation. To realize maximal symmetry, the $SO(5) \times SO(5)^\prime$ global symmetry of these two bulk fermions will be broken to $SO(5)_{V^\prime}$  by  the Yukawa coupling to the scalar $\Sigma$ mixing $\Psi$ and $\Psi'$ at the last site.  The general Lagrangian of this N-site model is of the form
\bea
&&\mathcal{L}_f =\bar{q}_L i\slashed D q_L+ \bar{t}_R i\slashed D t_R + \sum_{i=2}^N ( \bar{\Psi}_i i\slashed D_i \Psi_i ) \nonumber \\
&-&\epsilon_1 \bar{\Psi}_{q_L} U_1 \Psi_{2R}  -\sum_{j=2}^{N-1}  \epsilon_j\bar{\Psi}_{jL}   U_j \Psi_{j+1R}  - M_i \sum_{i=2}^N\bar{\Psi}_{iL} \Psi_{iR}   \nonumber \\
&-& \epsilon_1^\prime \bar{\Psi}_{t_R} U_1 \Psi'_{2 L}   -\sum_{j=2}^{N-1}  \epsilon_j^\prime \bar{\Psi'}_{jL}   U_j \Psi'_{j+1R}  - M_i^\prime \sum_{i=2}^N\bar{\Psi'}_{iL} \Psi'_{iR}   \nonumber \\
&-&  M  (\bar{\Psi}_{NL} \Sigma \Psi'_{NR} + \bar{\Psi'}_{NL} \Sigma \Psi_{NR}) +h.c.   
\eea
Note that due to the doubling of the bulk fermions all gauge invariant mass terms can be non-vanishing.  In this model, the global symmetry of the fermions at each intermediate site is $SO(5) \times SO(5)^\prime $, while at the last site we have the $SO(5)_{V^\prime}$. This symmetry breaking pattern is exactly what is needed to realize maximal symmetry. We can now explicitly find the low-energy effective Lagrangian for this model. First we redefine the phases of $\Psi'_i$ and $\Psi_i$ as in Eq.~(\ref{eq:redefinition}) and then re-do the integrating out of the bulk fermions site-by-site.  We again obtain the effective Lagrangian invariant under $SO(5)_{V^\prime}$
\bea
\mathcal{L}_f  &=& \bar{\Psi}_{q_L} \slashed p \Pi_0^L(p) \Psi_{q_L} + \bar{\Psi}_{t_R} \slashed p \Pi_0^R(p) \Psi_{t_R} \nonumber \\ 
&-& M_1(p) \bar{\Psi}_{q_L} U V U^\dagger  \Psi_{t_R} 
\eea 
confirming our expectation that maximal symmetry of the composite sector really forbids the Higgs dependent kinetic terms.   The key difference between the maximally symmetric N-site model and the ordinary deconstructed composite Higgs model is the unbroken global symmetry in the bulk fermion sector at the last site (in the ordinary moose, only the $SO(4)$ global symmetry is unbroken at the last site while here we have $SO(5)_{V'}$). More discussion about the ordinary deconstruction can be found in App.~\ref{app:MCHM}.

\section{Maximal Symmetry from Extra Dimensions}

The N-site model presented above can be directly generalized to a $5$D theory  warped in AdS space with a metric~\cite{Randall:1999ee}  
\bea
ds^2 =a(z)^2(\eta_\mn dx^\mu dx^\nu -dz^2) \equiv g_{MN} dx^M dx^N,
\eea 
where the coordinate $z$ along the fifth dimension ranges $R<z<R'$  and the warp factor $a(z) =\frac{R}{z}$ corresponds to an $S^1/Z_2$ orbifold with $R$ the AdS curvature. We will present the warped extra dimensional implementation of both the original maximally symmetric model as well as the new minimal maximal symmetry. 

\subsection{$\bf 5 +5$ bulk fermions - ordinary maximal symmetry}

We first show how to realize the original maximal symmetry in AdS space. From the construction of the N-site model we learn that for the  5D model using the $SO(5)/SO(4)$ MCHM coset~\cite{Agashe:2004rs}  we will need to introduce two separate bulk fermions $\Psi_1 $ and $\Psi_2$ in the fundamental representation of the $SO(5)$ bulk gauge symmetry. The zero modes of the $\Psi_1$ and $\Psi_2$ will be identified with the top doublet and singlet. To preserve an unbroken $SO(5)$ global symmetry at the IR brane, their boundary conditions (BCs) at the IR brane should be $SO(5)$ invariant. Hence the BCs will be

\bea \label{eq:bulk_fermion_bc}
\Psi_1
&=&  \!\left[\!\!
 \begin{array}{l}
   \mathbf{(2,2)^{q_1}_L} = \begin{bmatrix} q'_{1L}(-+) \\ q_{1L}(++) \end{bmatrix}    \\[0.5cm]
   \mathbf{(1,1)^{q_1}_L} = T_L (-+)   
 \end{array}  \!\!\right]\! , \;\; \nonumber \\
\Psi_2 &=& 
 \begin{bmatrix}
  \mathbf{(2,2)^{u}_R}(-+)   \\[0.15cm]
 \mathbf{(1,1)^{u}_R} = t_R (++)
 \end{bmatrix},
\eea
where we decomposed the $5$ dimensional vector representation of $SO(5)$ as $(2,2)+(1,1)$ under the unbroken $SO(4) \sim SU(2)_L\times SU(2)_R$ subgroup, 
$\pm$ denote Neumann or  Dirichlet BCs, and for $\Psi_1$ we only displayed the BCs for the left handed $\chi$ fields of the Dirac fermion (while for $\Psi_2$ the BCs are for the right handed fields). The fields with opposite chiralities will have opposite BCs. If these two bulk fermions do not mix with each other (ie. both have their respective bulk masses $m_{1,2}$ but no bulk mixing mass) their global symmetry at the IR brane will be $G_L \times G_R$:
\bea
\Psi_1 \to g_L \Psi_1 \quad \Psi_2 \to g_R \Psi_2 \quad g_{L,R} \in G_{L,R}. 
\eea

Since we chose the BCs  of the fermions at the IR brane to be $SO(5)$ invariant, the global $SO(5)_L \times SO(5)_R$ symmetry (which contains the Wilson line shift symmetry) will ensure that  the bulk gauge interactions do not contribute to a potential for the Wilson line $A_5$. In order to break the $A_5$ shift symmetry while still preserving the global $SO(5)_{co^\prime}$ symmetry we introduce the brane localized mass term mixing the two bulk fermions twisted by the Higgs parity  $V$:
\bea
S_{mix} =\frac{1}{g_5^2} \int d^4 x \sqrt{-g_{ind}} \tilde{m} ( \bar{\Psi}_{1L} V \Psi_{2R} +h.c.),   
\label{eq:boundarymix}
\eea
where $g_{ind}  $ is the induced metric at the IR brane. 
Due to the maximal symmetry, the Wilson line can be removed from the fermion sector if we turn off the interactions of either $\Psi_1$ or $\Psi_2$. Hence (as suggested by the N-site model) one needs the KK modes of  the bulk fermions to connect the top doublet and top singlet with the Wilson line, while the effective kinetic terms of the SM fermions will be independent of the Wilson line. As expected, the Higgs potential can only depend on the effective top Yukawa coupling.  

Now we can explicitly calculate the holographic effective Lagrangian at the UV brane to verify our conclusions. We will choose $\chi_L=\Psi_{1L} (x,R)$ and $\psi_R =\Psi_{2R}(x,R)$ as the  holographic fields. Imposing the bulk equations of motion one can show that the entire holographic effective action becomes a boundary term  
\bea
S_4 =\frac{1}{2 g_5^2} \int \left[ d^4 x \sqrt{-g_{ind}}     (\bar{\Psi}_{1L} \Psi_{1R}  -\bar{\Psi}_{2L} \Psi_{2R} + h.c. )\right]_R^{R'},\nonumber \\
\eea  
 Note that the sign of the holographic action changes depending on whether the LH or the RH field is taken as the basic holographic field.

\begin{figure}
\begin{center}
\includegraphics[width=0.45\columnwidth]{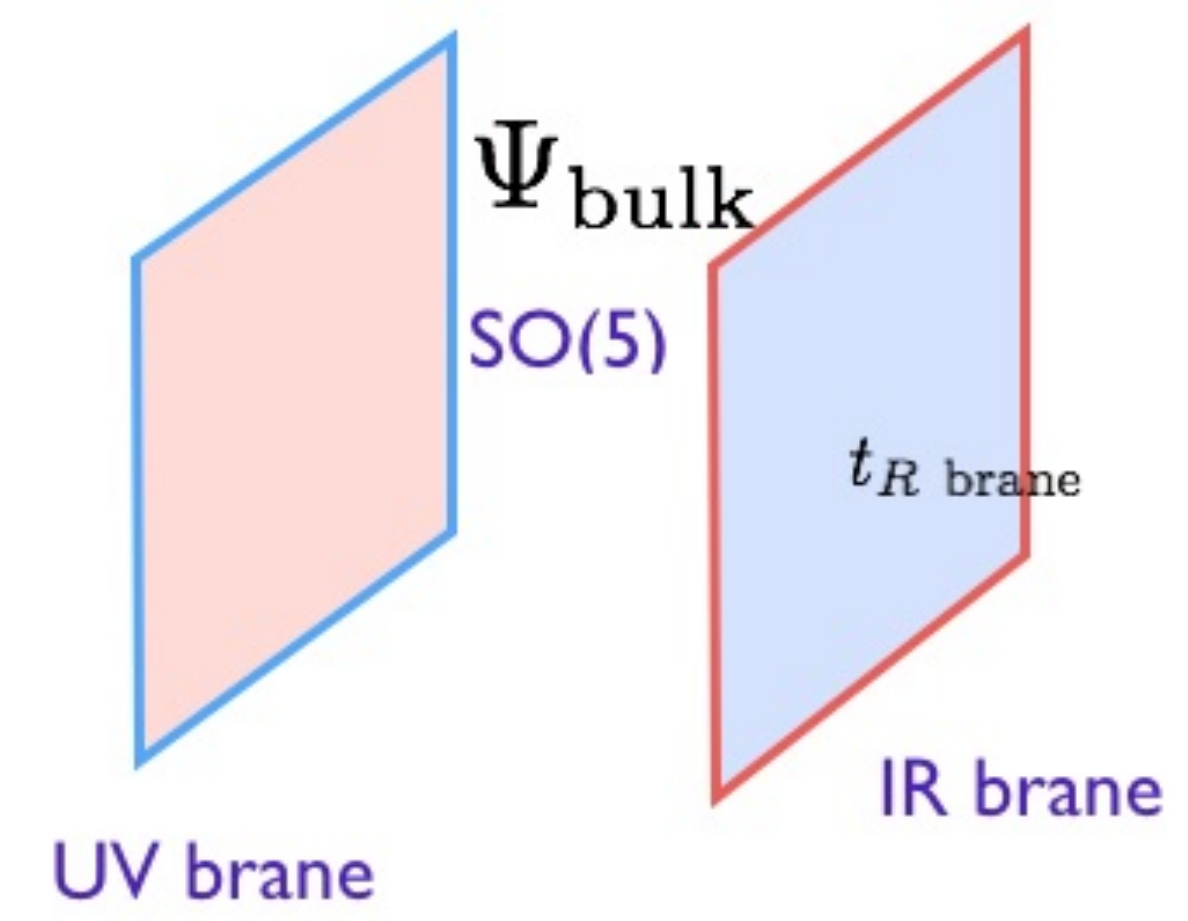}
\includegraphics[width=0.45\columnwidth]{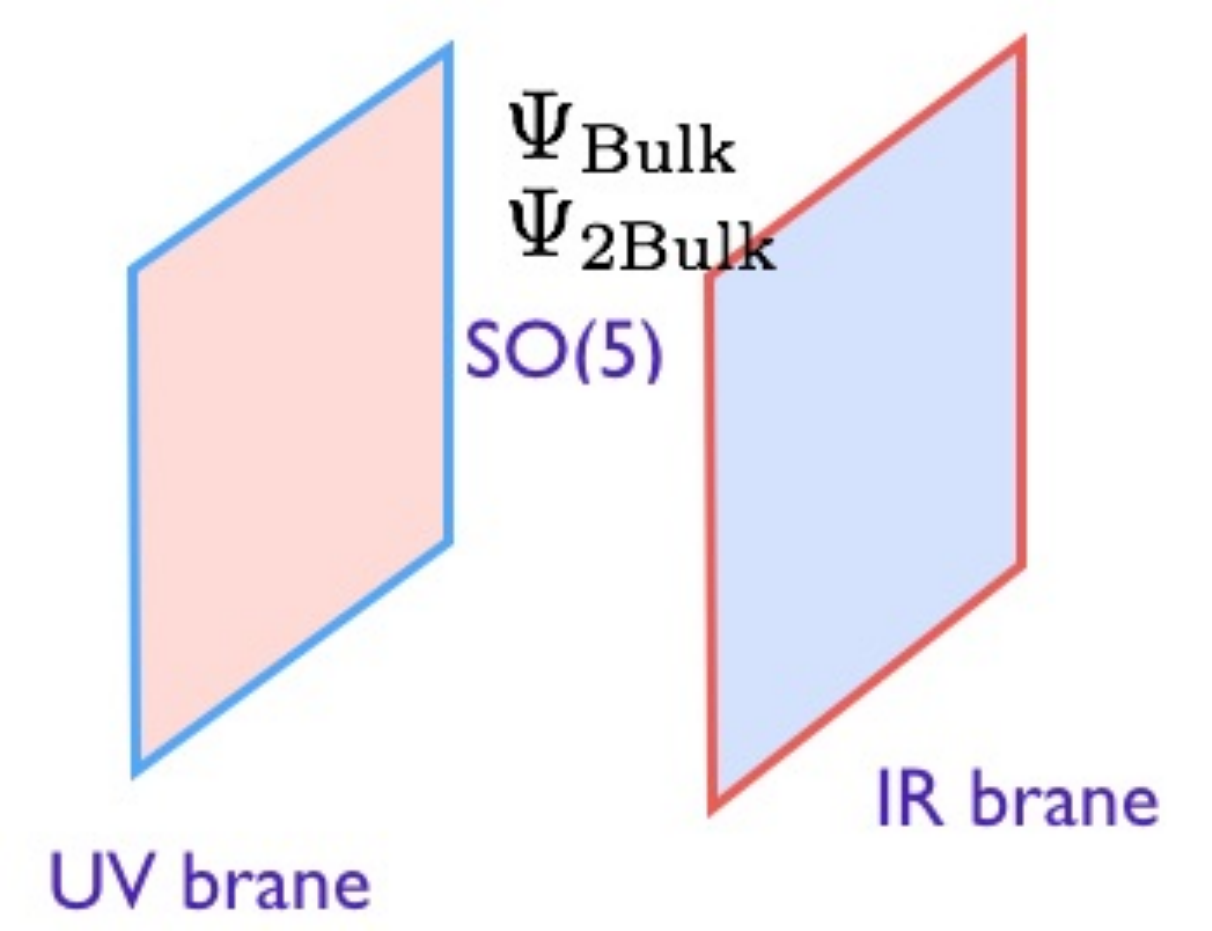}
\end{center}
\caption{Extra dimensional setup.} \label{fig:nsite3}
\end{figure} 
The addition of (\ref{eq:boundarymix}) will modify the IR boundary conditions of the fermions at $z=R'$ to 
\bea \label{eq:bc}
\Psi_{1R} =\tilde{m} V \Psi_{2R}  \quad \Psi_{2L} =-\tilde{m} V\Psi_{1L}. 
\eea 
Plugging the classical solutions of the bulk equations satisfying the BCs in Eq.~(\ref{eq:bc}), the holographic Lagrangian will be
\bea \label{eq:holo_55}
\mathcal{L}_{H} &=& \bar{\chi}_L \slashed p \Pi^L(\tilde{m}) \chi_L - \bar{\psi}_R \slashed p \Pi^R(\tilde{m}) \psi_R  \nonumber \\
&+&M^{LR} (\bar{\chi}_L V \psi_R +\bar{\psi}_R V\chi_L), \nonumber \\
\eea 
where the form factors $\Pi^{L,R}$ and $M^{LR}$ are explicitly shown in Eq.~(\ref{eq:form_factor_1}) in App.~\ref{app:form_factor}.

The presence of the non-vanishing Wilson line can be incorporated by a bulk rotation of the fields such that the Wilson line disappears from the bulk equations of motion, however will reappear in the BCs. The effect of this rotation will be the dressing of the holographic fields with the Wilson line via the expression~\cite{Contino:2010rs}
\bea
\chi_L,\psi_R &\to& U(R,R') \chi_L' ,\psi_R', \nonumber \\
  U(R,R') &=&  \B{Exp}\Big(i\frac{- \sqrt{2}\pi^{\hat{a}} T^{\hat{a}}}{f}  \Big),
\eea
where $f=\frac{2\sqrt{R}}{g_5 R'}$, $\pi^{\hat{a}}$ are the $A_5$ zero modes, and $T^{\hat{a}}$ is the generator in $SO(5)/SO(4)$. Finally the effective Lagrangian takes the form 
\bea
\mathcal{L}_{eff} &=&  \bar{\chi}_L^\prime  \slashed p \Pi^L(\tilde{m})   \chi_L^\prime -\bar{\psi}_R^\prime  \slashed p \Pi^R(\tilde{m}) \psi_R^\prime  \nonumber \\
&+&M^{LR} (\bar{\chi}_L^{\prime} \Sigma \psi_R^\prime +h.c.)    
\eea
where $\Sigma = U^\dagger V U =U^{\dagger 2} V$ and   
\bea \label{eq:holographic}
  \chi_L^\prime =\frac{1}{\sqrt{2}} \left( \begin{array}{c}
i b_L \\ 
b_L \\ 
i t_L \\ 
- t_L \\ 
0
\end{array} \right) \quad \psi_R^\prime = \left( \begin{array}{c}
0 \\ 
0 \\ 
0 \\ 
0\\ 
t_R
\end{array} \right). 
\eea  
We again see that due to the $SO(5)_{co^\prime}$ symmetry of the bulk sector, the effective kinetic terms are independent of $\Sigma$. Only the effective Yukawa coupling can depend on the NGBs, which is $SO(5)_{co^\prime}$ invariant. While the global symmetry is broken to $SO(4)$ by the $\Sigma$ VEV, the presence of  maximal symmetry in the top sector will again soften the divergences   consistent with our expectations from the N-site model.

\subsection{ $\bf 5 +1$ case: minimal maximal symmetry}

Finally we show the extra dimensional implementation of minimal maximal symmetry. Based on the structure of the N-site model corresponding to this scenario we expect that the  SM top doublet would still be a zero mode of the bulk fermion $\Psi_1$ in Eq.~(\ref{eq:bulk_fermion_bc}). However the top singlet $t_R$ should be a bulk singlet localized on the IR brane. The simplest choice is to add $t_R$ simply as a brane localized chiral Weyl fermion, though we can really think of it as the limit of a very heavy bulk fermion whose zero mode is sharply localized on the IR brane. 

 To realize minimal maximal symmetry, the bulk fermion at the IR brane should respect the $SO(5)_L$ global symmetry so its BC should be the same as the one in the previous case. Hence as in the previous analysis, the bulk fermion does not break the Wilson line shift symmetry.  To break the shift symmetry and produce the top Yukawa coupling, the singlet $t_R$ should interact with the Wilson line via the mixing with the bulk fermion. Since the mixing breaks the $SO(5)_L$ global symmetry to $SO(4)$ on the IR brane, the induced potential for the Wilson line must be proportional to the mixing parameter. Thus the $SO(5)_L$ global symmetry of the bulk fermion implies that the $A_5$ potential only depends on the effective top Yukawa coupling.    
Since the gauge symmetry at the IR brane is reduced to $SO(4)$, the $SO(4)$ singlet component of $\Psi_1$ can mix with the brane localized $t_R$:
\bea
S_{IR} &=& \frac{1}{g_5^2}\int d^4 x \sqrt{-g_{ind} }\Big(  \tilde{m} \bar{T}_L t_R + h.c. +\bar{t}_R \slashed D t_R \Big)  \nonumber \\ 
 \eea 
where $\slashed D =e^\mu_a \Gamma^a( \partial_\mu -i A_\mu )$. 
We can again choose the left-handed $\Psi_{1L}$ as the holographic field $\chi_L =\Psi_L^I(x, R)$. Again the holographic action will be a boundary term 
\bea
S_4 &=&\frac{1}{2 g_5^2} \int  d^4 x \sqrt{-g_{ind}}     \Big[\bar{\Psi}_{1L} \Psi_{1R} + h.c. \Big]_R^{R'},
\eea  
The boundary terms on the IR  brane will modify the BCs to be
 
\bea \label{eq:bc_IR}
T_{R}(x, R') =\tilde{m}  t_{R}  
\eea
Using this BC we can again obtain the classical solution in the bulk (for details see  App.~\ref{app:form_factor}).   
Following the same procedure as in the previous section (and normalizing the $t_R$ kinematic term by rescaling $t_R \to t_R (\frac{R'}{R})^\frac{3}{2}$, we can get the holographic effective Lagrangian  
\bea \label{eq:holo_51}
\mathcal{L}_H = \bar{\chi}_L \slashed p \Pi_L(p)   \chi_L +\bar{t}_R \slashed p t_R  +    \left(M(p) \bar{\chi}_L   t_R(p)+h.c.\right),  \nonumber \\ 
\eea  
where  the form factors $\Pi_L$ and $M$ are shown in detail in (\ref{eq:51_form}).    
Rotating the holographic field will restore the dependence on the NGBs fields, and turning off the component vanishing at the UV boundary,  
\bea
\chi_{L,R} \to U(R, R^\prime) \chi_{L}^\prime  \quad  \chi_L^\prime =\frac{1}{\sqrt{2}} \left( \begin{array}{c}
i b_L \\ 
b_L \\ 
i t_L \\ 
- t_L \\ 
0
\end{array} \right),
\eea
we obtain the effective Lagrangian for the  top sector
\bea
\mathcal{L}_{eff} &=&  \bar{\chi}_L^\prime \slashed p\Pi_L(p) \chi_L^\prime +\bar{t}_R \slashed p  t_R \nonumber \\
&+&\left( M(p) \bar{\chi}_L^{\prime} \mathcal{H} t_R +h.c. \right)    
\eea
where $\mathcal{H} =U^\dagger \mathcal{V}$ with $\mathcal{V} =(0,0,0,0,1)$.
As discussed before, the interval between the top doublet and singlet is $SO(5)$ invariant so their effective kinetic terms should be independent of the NGBs to preserve the global $SO(5)$, leading to the minimal maximal symmetry. 
\section{Comments on the Gauge Sector and the Higgs potential}

Maximal symmetry is the unbroken global symmetry $G$ in the (composite/bulk) fermion sector. However, in the gauge sector, there can not be such an unbroken  global symmetry because the global symmetry of the bulk gauge sector is the same as the gauge symmetry, which eventually must be broken to $H$. This statement can also be verified by inspecting the gauge boson effective Lagrangian which can be parametrized as 
\bea
\mathcal{L} = \frac{P_t^{\mu \nu}}{2} \Big( \Pi_0(p) \text{Tr}[A_\mu A_\nu ]   + \Pi_1(p) \text{Tr}[\Sigma A_\mu \Sigma A_\nu ]  \Big)
\eea              

The effects of the global symmetry in the composite sector can be traced by the dressed operator $U^\dagger A_\mu U$.  If there was a global symmetry $G$ in the composite gauge sector then the 
dressed operator would transform as $U^\dagger A_\mu U \to g U^\dagger A_\mu U g^\dagger$. We can see that while the $\Pi_0$ kinetic term is $G$ invariant (but independent of the NGB's), however the $\Pi_1$ term breaks $G$ to $H$ due to the VEV of the Sigma field. If $\Pi_1$ is vanishing there would be no gauge contribution to the Higgs potential. As soon as $\Pi_1$ is non-zero there can be no unbroken global symmetry larger than $H$. This proves that maximal symmetry can not be implemented in the composite gauge sector.  However the Higgs potential from the gauge sector can be still finite if the breaking of the global symmetries in the gauge sector is sufficiently collective (which is usually the case in deconstructed models with more than two sites).

Finally we explain the utility of maximal symmetry in achieving a phenomenologically viable Higgs potential. It is usually parametrized using the variable  $s_h \equiv \sin  \frac{\langle h \rangle}{f}  \ll 1$, and expanded to leading order as 
\bea 
V(h)= - (\gamma_f -\gamma_g)  s_h^2 +\beta_f s_h^4 
\eea  
where $\gamma_f, \beta_f$ are the contributions from the top sector and $\beta_g$ is the contribution from the gauge sector. If $\gamma_f -\gamma_g $ and $\beta_f$ are positive, the Higgs will acquire a VEV, $\xi \equiv  s_h^2 =  (\gamma_f -\gamma_g) /(2\beta_f)$. To achieve a  small $\xi$, the coefficient of the $s_h^2$ term has to be suppressed via cancellations.
The tuning measuring this cancellation is around 
\bea \label{eq:tuning}
\Delta \approx \frac{1}{\xi} \frac{\gamma_f}{  \beta_f  }. 
\eea

In a generic CHMs without any mechanism for softening the UV behavior of the Higgs potential, the leading contribution to $\gamma_f$ is  usually from the effective top kinetic terms and is quadratically divergent, while the leading contribution to  $\beta_f$  is log divergent, also originating from the top kinetic terms but at sub-leading order in $y_t$. Since these two terms have a different degree of divergence, the tuning in Eq.~(\ref{eq:tuning}) will usually be very large  and the Higgs very heavy. In composite Higgs models based on deconstruction or Holographic Higgs models  the Higgs potential is usually finite. However  $\gamma_f$ and $\beta_f$  are still from the leading and sub-leading contributions of the top effective kinetic terms so they will have a different dependence on the top Yukawa coupling $y_t$: $\gamma_f \sim \mathcal{O}(y_t)$ and $\beta_f \sim \mathcal{O}(y_t^2)$. For a viable Higgs potential one first needs to tune $\gamma_f$ to be the same order as $\beta_f$ and then tune it to be $\xi \beta_f$ to get small $\xi$, which results in a  double tuning. In this model the tuning is around $\Delta \sim  g_f^2/\xi $ for a light Higgs, where $g_f \equiv M_f/ f$  and $M_f$ is the top partner mass.    Note however that in these models the Higgs potential could be completely generated by the top sector.

In the CHM with maximal symmetry,  the Higgs potential from the top sector originates entirely from the top Yukawa coupling. This contribution is finite, and there is no freedom to impose a  cancellation purely within $\gamma_f$ to get a small $\xi$. The only way to achieve the suppression of the $s_h^2$ term is to use the contribution from the gauge sector $\gamma_g$   to cancel the contribution from the top sector $\gamma_f$. This is the main source of tuning   in  the maximally symmetric model. In the original maximally symmetric model~\cite{Csaki:2017cep} these two terms are of the same order in the  top Yukawa coupling thus the tuning is minimal, $\Delta \sim 1/\xi$.  

We want to emphasize that in all of the CHMs discussed so far in this section, the Higgs mass explicitly depends on the top partner mass, implying that the top partner mass must be light, around $g_f \approx 1$,  to obtain a $125$ GeV Higgs.    
However for the case of minimal maximal symmetry this situation changes.   
Due to the different choice of embeddings  the effective top Yukawa term is proportional to $s_h$ (vs. proportional to $s_{2h}$ in ordinary maximal symmetry), as a  result of which $\gamma_f\sim \mathcal{O}(y_t^2)$ and $\beta_f\sim \mathcal{O}(y_t^4)$.  
Hence in this scenario we still end up with a double tuning for small Higgs VEVs, similar to the  ordinary CHM. However $\beta_f$ is at  $\mathcal{O}(y_t^4)$ and can be parametrized as
\bea
\beta_f \approx c \frac{M_f^4}{(4\pi)^2 } (\frac{y_t}{g_f})^4,   
\eea
where $c$ is a numerical constant.
We can see that $\beta_f$  is not sensitive to  the top partner mass, which leads to an expression for the Higgs mass of the form $m_h^2  = 8\beta_f \xi /f^2\sim c/(2 \pi^2) y_t^4 v^2$.  We find  the Higgs is always somewhat too light, $m_h \approx 100$ GeV for $M_f \approx 10 f$. However the  Higgs mass can be easily enhanced while the tuning still significantly  suppressed if an independent Higgs quartic coupling can be produced through some mechanism~\cite{Ssaba:2018,6d}. For this case we will end up with a simple model where one only needs a reasonably small (smaller than for the traditional MCHM) tuning  to get a small $\xi$,  while at the same time  the Higgs mass is not sensitive to the top partner mass avoiding the LHC direct detection constraints.  For example fixing $\xi =0.1$ as well as the Higgs mass $m_h =125$ GeV we find that in this model the tuning is around $\Delta \approx 40$ if the lightest top partner is heavier than $2$ TeV.  On the other hand for the traditional composite Higgs models one needs a tuning larger than $100$ to keep the lightest top partner  heavier than $2$ TeV~\cite{Panico:2012uw}. In these models the masses of the Higgs and the top partners are strongly correlated and it is very difficult to obtain heavy top partners while keeping Higgs light. For these traditional models, one needs a careful tuning to obtain small $\xi$ and a separate tuning to keep the Higgs light, while for the model with minimal maximal symmetry the only tuning needed is to obtain small $\xi$.

\section{Conclusions and outlook}

The composite sector of composite Higgs models may have emergent (accidental) global symmetries. This maximal symmetry will have far-reaching consequences for the structure of the Higgs potential. It can forbid the Higgs dependence of the effective top kinetic terms 
 thereby softening its UV behavior. The particular forms of the maximally symmetric Higgs potential may also minimize the tuning needed or allow heavy top partners. We have explained the general conditions for the emergence of maximal symmetry as well as its major consequences. 

We have shown two major options for implementing maximal symmetry depending on the embeddings of the top doublet and singlet into the global symmetry.  We find that the essence of the emergence of maximal symmetry in a deconstructed or extra dimensional theory is to preserve a global $SO(5)$ symmetry  of the top partners at the last site/IR brane. We also showed that the structure of the gauge symmetries can enforce the emergence of maximal symmetry in the top sector.  In the ordinary maximal symmetric model, the Higgs potential parameters $\gamma_f$ and $\beta_f$ are almost equal so the tuning is always minimal. However in the new model with minimal maximal symmetry, the double tuning will still be present   because $\gamma_f$ and $\beta_f$ are order $\mathcal{O}(y_t^2)$ and $\mathcal{O}(y_t^4)$ in the top Yukawa coupling. The critical aspect of this new implementation of maximal symmetry is  that the Higgs mass is not sensitive to the top partner mass so a light Higgs can be obtained without light top partners.

\section*{Acknowledgements} 
 
  C.C.  thanks the Aspen Center for Physics - supported in part by NSF PHY-1607611 - and the ITP of the CAS in Beijing for its hospitality while working on this project. T.M. thanks the Cornell Particle Theory group for its hospitality while finishing this project. C.C. is supported in part by the NSF grant PHY-1719877 as well as the BSF grant 2016153. J.S. is supported by the NSFC under grant No.11647601, No.11690022, No.11675243 and No.11761141011 and also supported by the Strategic Priority Research Program of the Chinese Academy of Sciences under grant No.XDB21010200 and No.XDB23000000. T.M. is supported in part by project Y6Y2581B11 supported by 2016 National Postdoctoral Program for Innovative Talents. J.H.Y. is supported by the National Science Foundation of China under Grants No. 11875003 and the Chinese Academy of Sciences (CAS) Hundred-Talent Program.    

\appendix 
\section{Calculation of the  Holographic Form Factors}
\label{app:form_factor}

In AdS background,  the equation of motion for the bulk fermion field in mixed momentum-coordinate space $(p,z)$ are~\cite{Contino:2004vy}
\bea
(\partial_z+2 \frac{\partial_z a(z)}{a(z)} \pm a(z) m)\Psi_{L,R}  = \pm \slashed p \Psi_{R,L}
\eea
 If we parametrize the bulk fermion field as 
\bea
\Psi_{L, R}(p,z) =f_{L,R}(p,z) \Psi_{L,R}(p).
\eea
with $\slashed p \Psi_{L}(p) =p \Psi_{R}(p)$, we obtain the following equation 
\bea \label{eq:first_order}
(\partial_z+2 \frac{\partial_z a(z)}{a(z)} \pm a(z) m) f_{L}(p,z) =  \pm p f_{R}(p,z) 
\eea
Rewriting the equation of motion  for $f_{L,R}$ as
\bea \label{eq:second_order}
\Big( \partial_z^2 +p^2 -\frac{4}{z} \partial_z +\frac{6}{z^2}  \mp \frac{mR}{z^2}  - \frac{(m R)^2}{z^2} \Big) f_{L,R} =0 
\eea  
 we obtain the solutions
\bea
f_L(p,z) &=& z^{5/2} 
  \left[ J_\alpha(pz)Y_{\beta}(pR') - J_{\beta}(pR')Y_{\alpha}(pz) \right] \nonumber \\
f_R(p,z) &=& z^{5/2} 
  \left[ J_{\alpha-1}(pz)Y_{\beta}(pR') - J_{\beta}(pR')Y_{\alpha-1}(pz) \right], \nonumber \\ 
\eea
where $\alpha  \equiv m R +1/2$ and $\beta$ is still undetermined. 
To determine $\beta$, the IR BC is needed, which is different for the $5+5$ and $5+1$ cases. 

For our convenience, we define the following propagators
\bea
 G^{++}(\alpha_i) &\equiv f_L(p,R)\vert_{ \beta=\alpha_i } \quad 
 G^{-+}(\alpha_i) &\equiv f_R(p,R)\vert_{ \beta=\alpha_i }.\nonumber \\
 G^{+-}(\alpha_i) &\equiv f_L(p,R)\vert_{ \beta=\alpha_i-1 } \quad 
 G^{--}(\alpha_i) &\equiv f_R(p,R)\vert_{ \beta=\alpha_i -1},\nonumber \\
\eea
with $\alpha_i  \equiv m_i R+1/2$.

For $\bf 5+5$ case, the form factor is~\cite{Archer:2014qga,Croon:2015wba} 
\bea \label{eq:form_factor_1}
\Pi^{L} &=& \frac{1}{p}\frac{G^{--}(\alpha_1) G^{-+}(\alpha_2) - \tilde{m}^2 G^{-+}(\alpha_1) G^{--}(\alpha_2) }{G^{+-}(\alpha_1) G^{-+}(\alpha_2) + \tilde{m}^2 G^{++}(\alpha_1) G^{--}(\alpha_2)} \nonumber \\
\Pi^{R} &=& \frac{1}{p}\frac{G^{+-}(\alpha_1) G^{++}(\alpha_2) - \tilde{m}^2 G^{++}(\alpha_1) G^{+-}(\alpha_2) }{G^{+-}(\alpha_1) G^{-+}(\alpha_2) + \tilde{m}^2 G^{++}(\alpha_1) G^{--}(\alpha_2)}
\nonumber \\
 M^{LR} &=& \frac{\tilde{m}}{2} \frac{G^{-+}(\alpha_1) G^{+-}(\alpha_1) +  G^{++}(\alpha_1) G^{--}(\alpha_1)  }{G^{-+}(\alpha_1) G^{+-}(\alpha_2) + \tilde{m}^2 G^{++}(\alpha_1) G^{--}(\alpha_2)} \nonumber\\
 &+& \frac{\tilde{m}}{2} \frac{G^{-+}(\alpha_2) G^{+-}(\alpha_2) + G^{++}(\alpha_2) G^{--}(\alpha_2)}{G^{+-}(\alpha_1) G^{-+}(\alpha_2) + \tilde{m}^2 G^{++}(\alpha_1) G^{--}(\alpha_2)}.  \nonumber \\        
\eea

For the $\bf 5+1$ case, we have the solutions for the right-handed field at the UV brane~\cite{Archer:2014qga}
\bea \label{eq:51_solution}
T_R(p,L_0) &=& \frac{G^{--}(\alpha_1)}{G^{+-}(\alpha_1)} \frac{\slashed p}{p} \chi_L^5 +\frac{\tilde{G}^{+-}(\alpha_1)}{G^{+-}(\alpha_1)} \tilde{m} t_R(p), \nonumber \\
\Psi_{1R}^a(p,L_0) &=& \frac{G^{--}(\alpha_1)}{G^{+-}(\alpha_1)} \frac{\slashed p}{p} \chi_L^a, 
\eea
with 
$$\tilde{G}^{+-}(\alpha) = z^{5/2} \left[ J_\alpha(pR)Y_{\alpha-1}(pR) - J_{\alpha-1}(pR)Y_{\alpha}(pR) \right] $$ and index $a = 1, \cdots, 4$, 
and thus the form factors read
\bea \label{eq:51_form}
\Pi_L = \frac{1}{p}\frac{G^{--}(\alpha_1)  }{G^{+-}(\alpha_1)},  \quad M = \tilde{m} \frac{\tilde{G}^{+-}(\alpha_1)}{G^{+-}(\alpha_1)}. 
\eea

\section{Traditional composite Higgs  model in 4D moose}
\label{app:MCHM}

The essential difference between a traditional deconstructed MCHM and a model with maximal symmetry is 
whether the global $SO(5)$ symmetry of  the bulk fermions remains unbroken at the last site.
 In the traditional approach only the $SO(4)$ subgroup of $SO(5)$ at last site is gauged. The $SO(5)_L \times SO(5)_R$ global chiral symmetry  of the  top partners  is broken to the $SO(4)$ gauge symmetry at the last site. We will explain that this will lead to  the Higgs dependence of top effective kinetic terms. To simplify the discussion we only focus on the interactions  of the top doublet and its partners, which is given by       
\bea
&&\mathcal{L}_f =\bar{q}_L i\slashed D q_L + \sum_{i=2}^{N-1} ( \bar{\Psi}_i i\slashed D_i \Psi_i ) \nonumber \\
&-&\epsilon_1 \bar{\Psi}_{q_L} U_1 \Psi_{2R}  -\sum_{j=2}^{N-2}  \epsilon_j\bar{\Psi}_{jL}   U_j \Psi_{j+1R}  -\sum_{i=2}^{N-1} M_i \bar{\Psi}_{iL} \Psi_{iR}   \nonumber \\
&-& \epsilon_{N-1} \bar{\Psi}_{N-1 L} U_{N-1}( \Psi_{N R}^4 +\Psi^1_{N R}) \nonumber \\
&-&  M_4 \bar{\Psi}_{N R}^4 \Psi^4_{N L} -M_1 \bar{\Psi}_{N R}^1 \Psi^1_{N L} +h.c.,   
\eea     
where $D_i = \partial - i g_{\rho_i} \rho_i$ and $\rho_i$ and $g_{\rho_i}$ are the $SO(5)$ gauge bosons and gauge coupling at the $i^{th}$ site.    
Since the gauge symmetry at the last site is $SO(4)$, the $\Psi_N^{4,1}$ are forming an $SO(4)$ quadruplet and singlet. All the information of the $SO(5)/SO(4)$ breaking is encoded in the mass terms $M_4 \neq M_1$. In the extra dimensional case the corresponding statement is that the IR  boundary condition of the  bulk fermion corresponding to the top doublet is only $SO(4)$ invariant  (while their Yukawa couplings are assumed to be global $SO(5)$ invariant). 
Since the intermediate fields respect the global $SO(5)$ symmetry, integrating them out (as in the main text)   the contributions to the top kinetic terms remain independent of the Higgs  and only the effective Yukawa couplings of the top doublet $\Psi_{q_L}$ to $\Psi_N ^{4,1}$ can depend on the  non-linear sigma field $U \equiv \prod_{i=1}^{N-1} U_i $ containing the Higgs (which  corresponds to the Wilson line in the extra dimensional case). Since the fermion masses at the last site only respect $SO(4)$, which breaks the Higgs shift symmetry, after integrating out the Dirac fermions $\Psi_N^{4,1}$ at the last site the $\Psi_{q_L}$ effective kinetic term will in general pick up a dependence on the linearly realized pNGB field $\Sigma$. This reflects that the global symmetry of the doublet top partners is only $SO(4)$. We will find a similar conclusion in the extra dimensional warped MCHM: 
since the IR BCs of the bulk fermions corresponding to the top doublet and singlet only respects a global $SO(4)$   the effective top  kinetic terms in general will depend on the Wilson line.

 \end{document}